\pdfoutput=1
\documentclass[num-refs]{wiley-article}
\usepackage{graphicx}
\usepackage{siunitx}
\usepackage{amssymb}
\usepackage{latexsym}
\usepackage{amsfonts}
\usepackage{multicol}
\usepackage[utf8]{inputenc}
\papertype{Communication}
\paperfield{Journal Section}
\title{Understanding quality control of hard metals in industry - A quantum mechanics approach }
\author[1\authfn{1}]{\it Martina Lattemann}
\author[2\authfn{2}]{\it Ruiwen Xie}
\author[2\authfn{2}]{\it Raquel Lizárraga}
\author[2\authfn{2},3\authfn{3},4\authfn{4}]{\it Levente Vitos}
\author[1\authfn{1}]{\it Erik Holmstr{\"o}m}
\affil[1]{Sandvik Coromant R\&D, Stockholm SE-12680, Sweden}
\affil[2]{Applied Materials Physics, Department of Materials Science and Engineering, Royal Institute of Technology, Stockholm SE-10044, Sweden}
\affil[3]{Department of Physics and Astronomy, Division of MaterialsTheory, Uppsala University, Uppsala, Sweden}
\affil[4]{Wigner Research Centre for Physics, Institute for Solid State Physics and Optics, Budapest, Hungary}

\corraddress{Martina Lattemann, Sandvik Coromant R\&D, Stockholm SE-12680, Sweden}
\corremail{martina.lattemann@sandvik.com}
\presentadd[\authfn{1}]{Sandvik Coromant R\&D, Stockholm SE-12680, Sweden}
\fundinginfo{Project COFREE funded by European Institute of Innovation \& Technology (EIT), Grant Number. 15048; Ministry of Science and Technology (No.2014CB644001), Swedish Research Council; the Swedish Foundation for Strategic Research (SSF); Carl Tryggers Foundations; Swedish Foundation for International Cooperation in Research and Higher Education; Hungarian Scientific Research Fund, OTKA109570; China Scholarship Council} 
\runningauthor{Martina Lattemann et al.}
\begin{document}
\maketitle
\begin{abstract}
For many decades, the magnetic saturation of, e.g. hard metals (HM) such as WC-Co-based cemented carbides, has been used as process and quality control in industry to ensure consistency of product properties. In an urge of replacing cobalt as a binder phase, a demand on understanding the magnetic response as a function of composition on the atomic scale is growing. In this paper, a theoretical description of the measured weight specific magnetic saturation of hard metals as a function of the tungsten weight fraction present in the cobalt binder phase, based on first-principle calculations, has been established for standard WC-Co. The predicted magnetic saturation agrees well with the experimental one. Furthermore, it is proposed that the theoretical description can be extended to alternative and more complex binder phases which allows to transfer the production control to those hard metals.
\end{abstract}
Hard metals (HM) are widely used in industrial applications since 1920s due to their unique 
properties such as relative high hardness, toughness, and fatigue resistance. Hard metals consist
in their simplest and most common form of a hard phase, e.g. tungsten carbide (WC), and a ductile
binder phase, e.g. cobalt (Co). The properties are mainly dependent on their overall chemistry
which itself determines the fraction of tungsten (W) and carbon (C) dissolved in the Co binder
phase after sintering. The fraction of W and C solutes in the Co binder phase strongly affects the
grain growth and microstructure and, consequently, both the performance as well as the magnetic
properties of the hard metal. Therefore, magnetic saturation measurements are used in industry as a
fast, reliable and non-destructive method to assess consistency of properties and performance of
hard metals. More specifically, magnetic saturation measurement is a means of locating hard metals
within the so-called carbon window in the W-C-Co phase diagram, i.e. in which only the desired
phases exist. These measurements are not only used as quality control in production, but they are
equally crucial for the control of several production processes as well as in the development of
compositions of new hard metals.\\\newline
\noindent
However, a well-established relationship between the binder phase
composition and the overall properties and performance of hard metals is required to determine the location of samples within the carbon window. The accuracy needed for estimating the C content 
from the magnetic saturation measurements for samples prepared under closely comparable conditions is about 0.01~\%~\cite{upa98}. 
Roebuck et al.~\cite{roe84,roe96} assessed an empirical equation~\cite{freytag78} describing the relationship between the measured 
magnetic saturation of an insert and that of pure cobalt as a linear function of the tungsten 
weight fraction dissolved in the Co binder phase. 
In this empirical equation, it was assumed that the measured magnetic saturation is not affected by the dissolution of C in the Co binder phase \cite{roe84,roe96}. This assumption seems realistic since the solubility of C in Co is very limited~\cite{ishida91}. The success of the non-destructive measurement of the magnetic saturation and the reliability of the empirical equation by Roebuck for production control is based on decades of experience, the collection and comparison of a tremendous amount of experimental data including those from magnetic saturation measurements, materials analysis and performance testing. 
Gathering the required amount of information is both expensive and time consuming. Furthermore,
the empirical relationship is only valid for cemented carbides consisting of WC-Co without other
alloying elements and whenever a new alloying element is added to the Co binder, a new empirical
relation must be established, again based on experimental data collection. Due to the ever increasing requirements on quality and performance as well as on replacing Co as binder phase, a more fundamental understanding of the magnetic properties of hard metals is imperative in order to speed up the process of finding alternatives to Co as the binder phase. Thus, there is an urge to find a method to transfer the existing quality control to other binder phases. Recently, the current authors have developed such a method based on the approach presented in this work and applied it to a cemented carbide with FeNi binder~\cite{xie2018,linder2018}. \newline

\noindent
The aim of the presented work is to gain fundamental knowledge on the magnetic saturation of hard providing a theoretical framework of the empirical equation of Roebuck by using first-principles calculations.

\section{Theory}
\subsection{Quantum mechanics approach}
In the hard metal industry, the Cobalt magnetic saturation (Com), of a WC-Co hard metal insert is 
normally determined in accordance with the international standard (IEC 60404-14).
The weight specific magnetic saturation of the WC-Co hard metal insert (4$\pi\sigma_{HM}$) is measured and compared to the magnetic saturation per unit weight of pure Co (4$\pi\sigma_{Co}$). 

\noindent
Based on above description, Com of a WC-Co hard metal insert can be expressed as  
\begin{equation}
Com =\frac{4\pi \sigma_{HM}}{4\pi\sigma_{Co}}=\frac{\Big(\frac{M_{HM}}{m_{HM}}\Big)}{\Big(\frac{M_{Co}}{\hat{m}_{Co}}\Big)}
\label{eq_Com}
\end{equation}
where $M_{HM}$ is the total magnetic moment of the hard metal insert, $M_{Co}$ is the total magnetic moment of a pure Co reference material, $m_{HM}$ is the mass of the hard metal insert and $\hat{m}_{Co}$ is the mass of the pure Co reference sample.
Since $M_{HM}$ comes mainly from the Co binder phase. it is usualy close to $M_{Co}$ and, thus, the Com value of a hard metal insert is hence close to the weight fraction of Co in that hard metal insert. 
However, since non-magnetic W and C will be dissolved in the Co binder phase during the sinter process according to the W-Co-C phase diagram, the measured Com value will always be below the true weight fraction of Co in the hard metal insert.
Moreover, since the W concentration in the Co binder phase is closely related to the overall C weight concentration, Com provides an accurate measure of the total C concentration in the insert. 
This fact is used to relate Com values to the binder phase composition as well as the morphology and performance of hard metals. 
For comparison of hard metals with different Co binder contents, the relative weight specific magnetic saturation, Com/Co(wt.\%), is used. 
Co(wt.\%) is the intended weight fraction of Co binder phase in the hard metal insert and can be expressed as 
\begin{equation}
Co(wt.\%) = \frac{\hat{m}_{Co}}{m_{Ref}}
\label{eq_Co}
\end{equation}
where m$_{Ref}$ is the mass of the hard metal with pure Co as binder phase.
By using Equation~\ref{eq_Com} and \ref{eq_Co} Com/Co(wt.\%) becomes 
\begin{equation}
    \frac{Com}{Co(wt.\%)} = \frac{M_{HM}}{M_{Co}}\frac{m_{Ref}}{m_{HM}} = \frac{M_{HM}}{N_{Co}\cdot\mu_{Co}}\frac{m_{Ref}}{m_{HM}}
\label{ComCo_theo1}
\end{equation}
with N$_{Co}$ and $\mu_{Co}$ being the total number of Co atoms and the  magnetic moment of pure Co per Co atom, respectively.
Thus, Com/Co(wt.\%) can be understood as a measure of how much the weight specific magnetic moment of the cemented carbide is lowered by the dissolution of W and C in the Co binder phase. \newline 
For the theoretical description of the relative magnetic saturation of a hard metal, only the binder phase needs to be considered since WC is non-magnetic. 
We also follow the assumption by Roebuck et al.~\cite{roe84}, that the dissolution of C in the binder phase and its effect on the magnetic saturation is negligible so that the binder phase can be considered as a Co - W alloy. We further assume that the alloy is a random solid solution without clustering of W.
%
\vfill
\begin{figure}[hbt!]
    \centering
    \includegraphics[width=0.7\textwidth]{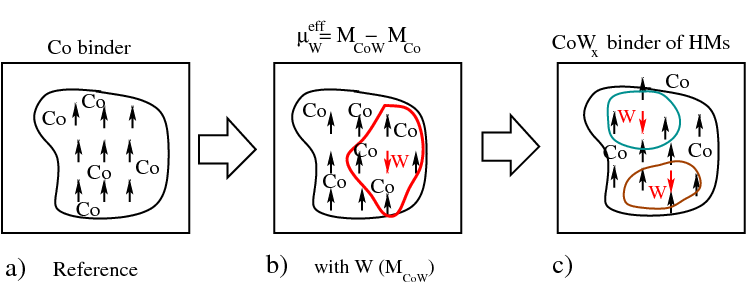}
    \caption{ Schematic of the theoretical approach to determine Com. (a) The magnetic moment of the pure Co binder is determined as the reference. (b) The magnetic moment of the dilutely alloyed Co binder phase is determined. The magnetic effect of a W impurity in Co is illustrated as a ''cloud'', describing the volume of Co binder with a changed magnetic moment. (c) Com is then determined by summing the appropriate amount of impurities for a certain W concentration.} 
    \label{fig1}
\end{figure}
The magnetic effect of the dissolved W in Co can be derived as schematically depicted in 
Figure~\ref{fig1}. 
Two calculations are required. Firstly, a reference calculation with pure Co, and secondly, a similar calculation in which one Co atom is changed to a W impurity. 
It is presumed that the magnetic moment of the Co 
atoms close to the dissolved W impurity is altered. 
This volume of Co atoms with altered magnetic moment 
is further denoted as a ''cloud'' (Figure~\ref{fig1}b). 
The total change of the magnetic moment in the Co binder
phase induced by a W solute is denoted as an effective W moment ($\mu^{eff}_W$). 
It is further assumed that the ''clouds'' around individual W solutes are not overlapping 
as seen in Figure~\ref{fig1}c. 
The total magnetic moment of the Co binder phase is then determined by the total amount of W solutes.
By using Equation~\ref{eq_Com} and \ref{eq_Co}, Com/Co(wt.\%) becomes 
\begin{equation}
    \frac{Com}{Co(wt.\%)} = \frac{N_{Co}\cdot\mu_{Co} + N_W^b\cdot\mu_{W}^{eff}}{N_{Co}\cdot\mu_{Co}}\frac{m_{Ref}}{m_{HM}}= \Big[1+\frac{N_W^b}{N_{Co}}\frac{\mu_W^{eff}}{\mu_{Co}}\Big]\frac{m_{Ref}}{m_{HM}}
\label{ComCo_theo2}
\end{equation}
where N$_W^b$ is the number of W atoms dissolved in the Co binder phase. In a first approximation, the mass of the hard metal sample measured can be written as m$_{HM}$~=~m$_{Ref}$~+~N$_W^b\cdot m_W$. To convert Equation~\ref{ComCo_theo2} into weight fraction of W, (w$_W^b$), dissolved in the Co binder phase, the following relation is used
\begin{equation}
w_W^b = \frac{N_W^b\cdot m_W}{N_{Co}\cdot m_{Co} + N_W^b\cdot m_W}
\label{attowt}
\end{equation}
where m$_W$ denote the atomic mass of W. 
By re-writing Equation~\ref{attowt} to replace N$_W^b$ divided by N$_{Co}$ in 
Equation~\ref{ComCo_theo2} we get
\begin{equation}
   \frac{Com}{Co(wt.\%)} = \Big(1 - \Big[1 + \frac{m_{Co}\cdot \mu_W^{eff}}{m_W\cdot\mu_{Co}}\Big]\cdot w_W^b\Big)\cdot\Big[\frac{1}{1-(1-Co(wt.\%))\cdot w_W^b}\Big]
\label{ComCo_final}
\end{equation}
If only the Co-W phase as alloy is considered as Roebuck et al.~\cite{roe84}, Equation~\ref{ComCo_final} is reduced to 
\begin{equation}
    \frac{Com}{Co(wt.\%)} =1 - \Big[1 - \frac{m_{Co}\cdot \mu_W^{eff}}{m_W\cdot\mu_{Co}}\Big]\cdot w_W^b
\label{ComCo_theo3}
\end{equation}
\begin{figure}[hbt!]
\centering 
\includegraphics[width=8cm]{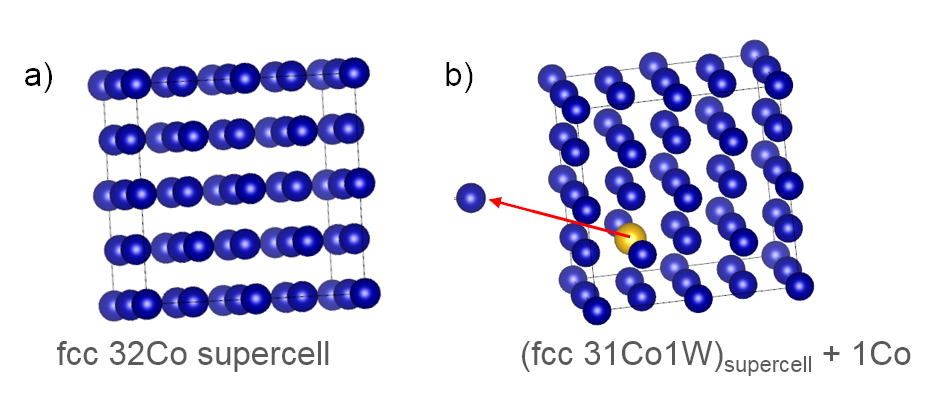}
\caption{Schematic of a 2~x~2~x~2 supercell structure consisting of 31 Co (blue) atoms plus one substitutional W (grey) solute atom.} 
\label{fig2}
\end{figure}
To determine the change of the magnetic moment in the Co binder phase induced per W solute, $\mu^{eff}_W$, 
first-principles calculations were performed by applying a supercell approach. 
In hard metals, the Co binder 
phase is normally stabilized in the metastable face-centered-cubic (fcc) structure (Figure~\ref{fig2}a) and 
due to the symmetry of the fcc structure, all Co atoms in the binder have the same 
magnetic moment. 
The total magnetic moment of the supercell, representing the binder, is given as 
{$M_{Co}=\sum_{i}\mu^{i}_{Co}$} with index i=\{1,N$_{Co}$\} and N$_{Co}$ is the total number of Co atoms 
in the supercell. 
To calculate the magnetic moment of the Co binder phase with one W solute (M$_{CoW}$), 
one of the Co atoms in the supercell is substituted by a W atom (see  Figure~\ref{fig2}b) and the change 
in the total magnetic moment of the supercell, i.e. the effective magnetic moment of a W solute in Co 
($\mu_W^{eff}$), is given by
\begin{equation}
    \mu_{W}^{eff}= (M_{CoW} + \mu_{Co}) - M_{Co}.
\label{muWeff}
\end{equation}
The additional magnetic moment $\mu_{Co}$ in the first term in Equation~\ref{muWeff} is the magnetic moment of the Co atom which was substituted by a W atom in order to keep the amount of Co binder atoms constant.
%
%
%
\subsection{First-principles calculations}
All first-principles calculations were performed using an all-electron projector-augmented wave (PAW) method 
as implemented in the Vienna Ab initio Simulation Package (VASP) code \cite{kre96,kre99,blo94}.
The generalized gradient approximation (GGA) \cite{per96} was used for treating electron 
exchange-correlation effects. A 2~x~2~x~2 fcc supercell comprising 32 atoms was used for the calculations.
For each supercell volume, the internal parameters were relaxed with a convergence criterion for the 
electronic subsystem 
to be equal to 10$^{-6}$~eV between two subsequent iterations, and the ionic relaxation loop within the 
conjugated gradient method was stopped according to its Hellmann-Feynman forces \cite{hel37,fey39} when 
the force on each ion was below 10$^{-3}$~eV/{\AA}. 
Brillouin zone sampling was performed using the Methfessel-Paxton
smearing method with SIGMA~=~0.2 \cite{met89} and a mesh of k-points centered at the $\Gamma$ point was 
carefully chosen to achieve entropy values below 1~meV. 
To compensate for the lattice 
expansion due to the substitutional W atom in the supercell, the lattice parameter was manually adjusted and 
the magnetic moment of the relaxed volume structure was determined. 
The energy cut-off for plane waves 
included in the expansion of wave functions was set to 1.3~ENMAX. All calculations were carried out at 
zero electronic and ionic temperatures. 
\begin{table}[h!]
    \centering
    \caption{Calculated magnetic moment of pure Co per Co atom ($\mu_{Co}$), induced magnetic moment of the W atom ($\mu_W$), the effective magnetic moment of a W solute ($\mu_W^{eff}$) using first-principles calculations and the atomic mass of Co and W.}
    \begin{tabular}{ c c c c c}
    \hline
        $\mu_{Co}$ [$\mu_B$] & $\mu_W$ [$\mu_B$] & $\mu_w^{eff}$ [$\mu_B$] & m$_{Co}$ [u] & m$_W$ [u] \\ 
       \hline\hline
         1.658 & -0.545 & -3.00 & 58.933 & 183.84 \\
         \hline
    \end{tabular}
    \label{tab1}
\end{table}
%
%
\section{Results and Discussion}
\begin{figure}[hbt]
\centering
    \includegraphics[width=8cm]{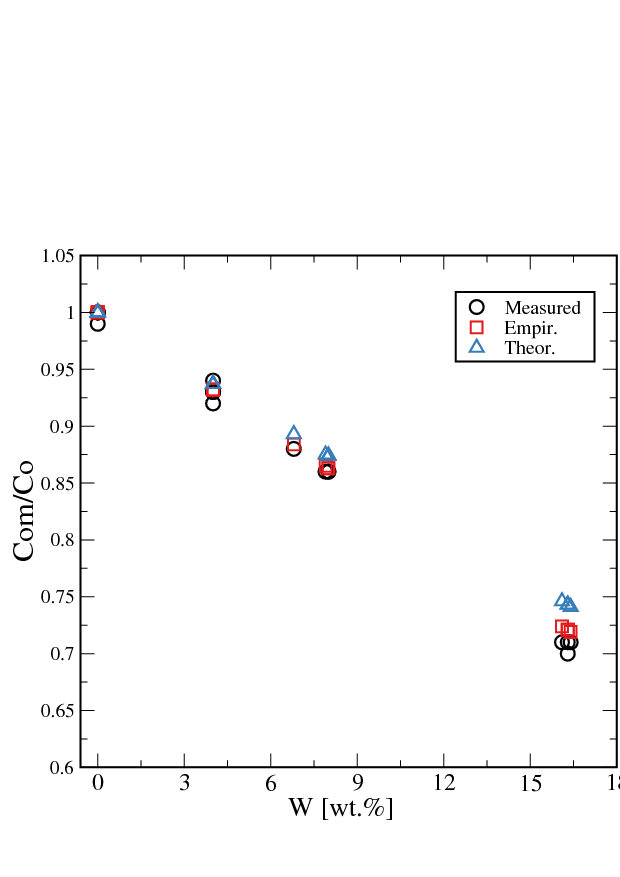}
    \caption{The calculated values for Com/Co(wt.\%) using Equations 9 and 11 are compared to the measured values published in Roebuck et al.~\cite{roe84}.}
    \label{fig3}
\end{figure}

To verify the applicability and accuracy of our model, the calculated Com/Co(wt.\%) are compared to 
the results found in hot extruded Co-W-C alloys~\cite{roe84}, i.e. no WC particles present 
(Co(wt.\%)~=~1).
From the above described first-principles calculations, the magnetic moment of Co per Co atom, $\mu_{Co}$, the 
induced magnetic moment of the W atom, $\mu_W$, and the effective magnetic moment of 
a W solute, $\mu_W^{eff}$, have been calculated and summarized in Table~\ref{tab1}.
\noindent
By applying the calculated values given in Table~\ref{tab1} and the atomic mass of Co and W to 
Equation~\ref{ComCo_theo3} the following theoretical expression for the change in magnetic saturation 
as function of dissolved tungsten in the Co binder is obtained
\begin{equation}
\left. \frac{Com}{Co(wt.\%)} \right\rvert_{Theor.} = ~~1 - 1.58\cdot w_W^b
    \label{final}
\end{equation}
The empirical equation for Com/Co(wt.\%) by Roebuck et al.~\cite{roe84,roe96} is given as
\begin{equation}
    4\pi\cdot\sigma_B = 4\pi\big[\sigma_{Co} - 0.275\cdot w_W^b\big] 
\label{eq_roe96}
\end{equation}
where $4\pi\sigma_{B}$ is defined as the measured magnetic saturation of a Co-W metal alloy per unit weight. 
By assuming the relation of Equation~\ref{eq_roe96} for the binder in a WC-Co cemented carbide we get 
\begin{equation}
 \left.\frac{Com}{Co(wt.\%)}\right\rvert_{Empir.} = ~~1 - 1.71\cdot w_W^b.
\label{ComCo_roe}
\end{equation}
%
%
For conventional WC-Co hard metals, i.e. not alloyed with other metals, Com/Co(wt.\%) are known to correspond to about 0.70 and 0.98 for the $\eta$-phase limit, i.e. at high W concentration dissolved in the Co binder phase, and the graphite limit, i.e. at low W concentration dissolved in the Co binder phase, respectively.
For this range of Com/Co(wt.\%), the concentration of W in the Co binder phase ranges from about 16.1~wt.\% and 1.4~wt.\% according to WC-Co phase diagram calculated using the Thermo-Calc software package~\cite{the02} and the parameters from~\cite{mark06}. 
Com/Co(wt.\%) values for these limits were calculated using Equations~\ref{final} and \ref{ComCo_roe} 
are presented in Table~\ref{tab2}. 
\begin{table}[htb]
    \centering
    \caption{Weight concentration of W dissolved in Co corresponding to carbon window, Com/Co(wt.\%) calculated using Equation~9 and 11.}
    \begin{tabular}{ c c c c}
    \hline
       W wt.\% & Theor. & Empir. & Measured\\ 
       \hline\hline
         1.4 & 0.978 & 0.976 & 0.93$\pm$0.01\\\
         16.3 & 0.742 & 0.721 & 0.705$\pm$0.005 \\
         \hline
    \end{tabular}
    \label{tab2}
\end{table}
In Figure~\ref{fig3}, the calculated values for Com/Co(wt.\%) using Equation~\ref{final} and \ref{ComCo_roe} together with the in Roebuck et al.~\cite{roe84} stated w$_W^b$ are compared to the therein published experimental Com/Co(wt.\%) values. \newline
It is verified that the calculated COM/Co(wt.\%) based on our theoretical model are in good agreement with both experimental values and Roebuck's empirical relationship with only a minor discrepancy. 
This discrepancy is assumed to be due to our approximations in the model for the first-principles calculations, namely, neglecting the effect of dissolved carbon, the possible presence of hcp Co in the Co binder phase is disregarded, the effect of WC-Co interfaces on the magnetic properties, and that a random solution of Co and W without overlapping magnetic clouds is assumed.
%

To investigate the correctness of the assumption that the clouds are not overlapping, the volume of the ''cloud'' of a W solute, i.e. number of affected Co atoms in the supercell, has been determined by comparing the magnetic moment of the Co atoms surrounding the W solute to the magnetic moment of pure Co per Co atom, counting all Co atoms with a changed magnetic moment of more than 0.08~$\mu_B$. 
The ''cloud'' of a W solute was then determined to comprise the 12 nearest neighbour atoms, i.e. 1 out of 13 atoms. 
As a result, an absolute maximum for the approximation of non-overlapping ''clouds'' to be is 7~at.\%, corresponding to about 19~wt.\%, which is above the upper limit of the W concentration in the Co binder phase, i.e. the $\eta$-phase limit of the carbon window. However, the random distribution of W atoms in the alloy will lower this limit and possibly contribute to the difference between theory and measurement that we can see at high W concentrations in Figure~\ref{fig3}. 
In the experimental results from Roebuck, the alloys had a varying grain size and morphology as function of W 
content. Such variations may also influence the magnetic saturation measurements and there was no such effect considered in the calculations. 

In the case of cemented carbide, the other main effect that will change the calculated Com/Co is the
magnetic environments at the WC-Co interfaces. 
In a recent investigation, this effect was estimated for a FeNi based binder alloy and it was 
found that the magnetic moments are lowered at the interfaces~\cite{xie2018}.
%
\section{Conclusions}
First-principle calculations of the magnetic moment of Co and a random Co-W solid solution alloys 
have been used to 
develop a theoretical model to calculate Com/Co(wt.\%) as a function of W wt.\% that can be applied to cemented carbide.
The comparison of experimental values and values determined using Roebuck's empirical equation, 
verifies that 
our approximations of non-overlapping ''clouds'' in a random solid solution of Co-W neglecting the effect of C 
dissolved is acceptable for the given concentration range of W and C in the Co binder phase. 
Therefore, our results provide reliable and sufficiently accurate prediction of Com/Co(wt.\%) that 
can be used in 
production control. 
The theoretical description of Com/Co(wt.\%) can be expanded to other materials and phases  
by performing similar calculations that describes the binder phase. 
In particular, the generalization from Co-based binders to alternative binders of alloys, e.g. 
FeNi-based binders, 
is possible due to the predictive power of first-principles calculations~\cite{xie2018}.
\section*{acknowledgements}
The present work is performed under the project COFREE (proj. no. 15048), funded by European Institute of Innovation \& Technology (EIT). The authors acknowledge the Ministry of Science and Technology (No.2014CB644001), the Swedish Research Council, the Swedish Foundation for Strategic Research (SSF), the Carl Tryggers Foundations, the Swedish Foundation for International Cooperation in Research and Higher Education, the Hungarian Scientific Research Fund (OTKA 109570), and the China Scholarship Council for financial supports.
The computations were performed on resources provided by SNIC through Uppsala Multidisciplinary Center for Advanced Computational Science (UPPMAX).


\bibliography{Com_manus}

\end{document}